\documentclass[fleqn,twoside]{article}
\usepackage{espcrc2}

\usepackage{graphicx}
\usepackage{epsfig}
\newcommand{\AmS}{{\protect\the\textfont2
  A\kern-.1667em\lower.5ex\hbox{M}\kern-.125emS}}

\title{
\vspace{-35mm}
\rightline{\small RNCP-Th01023~~~~~}
\rightline{\small UNITU-THEP 31/2001~~~~~}
\rightline{\small 11 October, 2001~~~~~}
\vspace{10mm}
Singular gauge potentials and the gluon condensate \\
              at zero temperature}

\author{K. Langfeld\address[ITPUT]{Institute for Theoretical Physics,
University of T\"ubingen, Germany}, 
        E.-M. Ilgenfritz\addressmark[ITPUT]\address[RCNP]{Research 
        Center for Nuclear Physics, Osaka University, Osaka 567-0047, 
        Japan}\thanks{Poster presented by E.-M.I.}\thanks{E.-M.I. 
        gratefully appreciates the support by the Ministry 
        of Education, Culture and Science of Japan (Monbu-Kagaku-sho) and
        thanks for a CERN visitorship.},
        H. Reinhardt\addressmark[ITPUT]
        and
        A. Sch\"afke\addressmark[ITPUT]
        }
       
\begin{document}

\begin{abstract}
We consider a new cooling procedure which separates gluon degrees 
of freedom from singular center vortices in $SU(2)$ LGT in a gauge
invariant way.
Restricted by a cooling scale $\kappa^4/\sigma^2$ fixing the residual $SO(3)$ 
gluonic action relative to the string tension, the procedure is RG invariant.
In the limit $\kappa \to 0$ a pure $Z(2)$ vortex texture is left. 
This {\it minimal} vortex content does not contribute to the string 
tension. It reproduces, however, the lowest glueball states.
With an action density scaling like $a^4$ with $\beta$, it defines 
a finite contribution to the action density at $T=0$ in the continuum 
limit. We propose to interpret this a mass dimension 4 condensate 
related to the gluon condensate. Similarly, this vortex texture is 
revealed in the Landau gauge.
 
\vspace{1pc}
\end{abstract}

\maketitle

\section{COSET COOLING}
$SU(N)$ gluodynamics at zero temperature is believed to confine since 
extended vortex degrees of freedom carrying $Z(N)$ flux are realized 
in a condensed phase. In LGT one attempts to localize, configuration 
by configuration, vortices by center projection, mostly starting from 
the Maximal Center Gauge (MCG) thought to be optimal for that purpose\cite{MCG}. 
The result, however, $Z(N)$ flux squeezed into thin $P$-vortices, is 
irreproducible with respect to the gauge copy to which MCG 
fixing is applied\cite{Drama}. In particular, starting from Landau gauge 
no vortex structure is discovered by subsequent MCG fixing which would 
allow to understand the area law of Wilson loops\cite{Kovacs}. 

A gauge independent method of vortex finding is therefore highly desirable.
This was the motivation to explore the capability of coset cooling\cite{NPB} 
for this purpose, complementing other methods making use of the lowest 
modes of some auxiliary adjoint Higgs field (Laplacian Center Gauge\cite{PdF}). 
Here, we will 
report that - instead of fixing the {\it confining} 
center vortices - the coset cooling method is rather localizing a minimal 
center vortex content living at the UV scale. It corresponds to singular fields 
which carry a finite action density in the continuum limit.
This disagrees with the belief that, approaching the continuum limit, 
the lattice degrees of freedom can be reduced to gluon fields 
$A_{\mu}$ by expanding $U_{x,\mu}=\exp \left(i a A^a_{mu}(x) t^a \right)$ 
around the unit element.
The same observation, based on a new implementation of Landau gauge fixing 
at zero and non-zero temperature, is reported by K.~Langfeld\cite{KL-talk}.

Coset cooling is relaxation with respect to the {\it adjoint} action which 
is written for a given link
\begin{equation}
s^{\mathrm{gluon}}_{x,\mu} = \frac{4}{3}~\sum_{\bar{\nu}=\pm1,\ne\pm \mu}^{\pm4}
\left( 1 - \left( \frac{1}{2}~\mathrm{tr}~U_{x,\mu \bar{\nu}} \right)^2  \right) .
\end{equation}
This action tolerates $Z(2)$ vortex degrees of freedom on the plaquette scale. 
The relaxation update consists in replacing $U_{x,\mu}$ 
by $U^{\mathrm cool}_{x,\mu}$ proportional to 
\begin{equation}
\sum_{\bar{\nu}=\pm1,\ne\pm \mu}^{\pm4} 
~\mathrm{tr}~( U_{x,\mu \bar{\nu}} )~\, 
U_{x,\bar{\nu}}~U_{x+\hat{\bar{\nu}},\mu}~U^{\dagger}_{x+\hat{\mu},\bar{\nu}} \, .
\end{equation}
We have used {\it restricted} coset cooling: when the local gluon 
action obeys $s^{\mathrm{gluon}}_{x,\mu} < 8~\kappa^4~a^a$, relaxation 
of the given link stops. Considering Monte Carlo ensembles generated at various 
$\beta$ values, the result of this cooling is RG invariant if and only 
if the cooling scale $\kappa/\sqrt{\sigma}$ is chosen independently of 
$\beta$. In the limit $\kappa \to 0$, coset cooling leads to a pure $Z(2)$ 
configuration. Otherwise, at any finite $\kappa$ (monitoring the residual 
gluon action density), the $Z(2)$ degrees of freedom are defined by 
usual center projection 
$U_{x,\mu} \rightarrow Z_{x,\mu} = \mathrm{sign}~\mathrm{tr}~(U_{x,\mu})$.  

\section{THE STATIC $Q\bar{Q}$ POTENTIAL IN COSET-COOLED LATTICE FIELDS}

We consider coset-cooled lattices representing Monte Carlo ensembles for 
different $\beta$ as a function of the cooling scale.
On one hand, for $\kappa/\sqrt{\sigma} \approx 1$ the Wilson action is already 
strongly concentrated on a subset of plaquettes.
On the other hand, at this stage coset-cooled configurations still sustain 
the full $Q\bar{Q}$ force over distances $R > 0.4$ fm. The short-range Coulomb 
force due to gluon exchange is already wiped out while extended vortices, 
which are relevant for the formation of color-electric flux tubes, are still 
intact.

\begin{figure}[!htb]
\vspace{-5mm}
\epsfxsize=6.6cm \epsffile{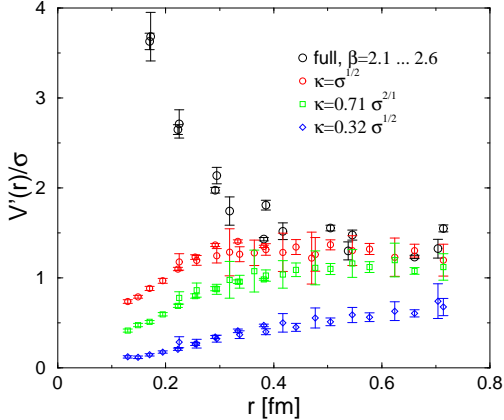}
\vspace{-12mm}
\caption
{Static $Q\bar{Q}$ force vs. distance $r$ for uncooled $SU(2)$ configurations
compared with the force for coset-cooled configurations} 
\label{fig:1}
\end{figure}
\vspace{-8mm}

Fig. 1 shows for a $12^4$ lattice how the $Q\bar{Q}$ force measured on 
coset-cooled configurations changes with decreasing $\kappa/\sqrt{\sigma} \leq 1$. 
Data corresponding to various $\beta$ values fall on curves characterized 
by the cooling scale. Going to smaller $\kappa/\sqrt{\sigma}$, the delayed onset
of the string tension can be explained by the suppression of extended vortices 
with a thickness below the cooling radius. This is convincing, at least for 
the cooling scales 1 and 0.71.

Thus the proposed cooling {\it does not preserve} the non-perturbative 
force at all distances. Instead, we have found a method to suppress 
thick confining vortices below some cooling radius. This might become useful 
in future studies of the confinement mechanism. In the following we will
rather concentrate on the singular vortex component ($c$-vortices) 
disclosed by the cooling technique. 

\section{THE VORTEX TEXTURE AS A $d=4$ VACUUM CONDENSATE}
 
Consider the trace of the energy-momentum tensor 
$\theta^{\mu}_{\mu} $ 
which appears as the dimension four vacuum condensate $O_4$ in 
the operator product expansion. 
We found that $O_4~a^4$ scales like $a^4$ with $\beta > 2.2$ for 
different values of the cooling scale $\kappa/\sqrt{\sigma}$.

Fig. 2 shows for two values of $\kappa/\sqrt{\sigma}$ the dimensionless lattice 
condensate
\begin{equation}
O_4(\kappa)~a^4 = 
\frac{24}{\pi^2}~\langle 1~-~\frac{1}{2}~\mathrm{tr}~U_p 
\rangle_{|\mathrm{cooled~with~scale~}\kappa}
\end{equation}
as a function of $\beta$. Already for finite $\kappa$ we find  
\begin{figure}[!htb]
\vspace{-5mm}
\epsfxsize=6.6cm \epsffile{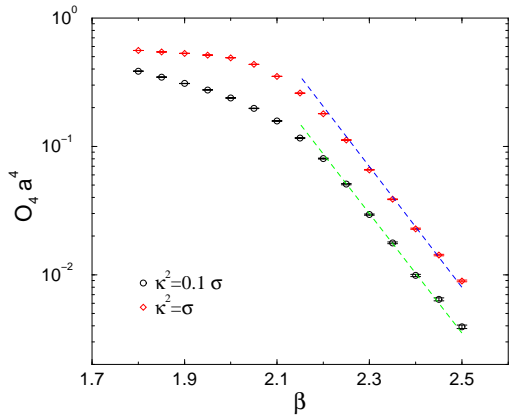}
\vspace{-12mm}
\caption
{$O_4$ in lattice units for two different coset cooling scales vs. $\beta$}
\label{fig:2}
\end{figure}
\vspace{-6mm}
the condensate $O_4(\kappa)$ as a well defined, RG invariant function of $\kappa$
in the continuum limit. We have tried to describe this condensate as a function 
of $\kappa$ by a fit of the form $O_4(\kappa)/\sigma^2 = a_0 + a_1 \kappa^4$
with (fit A) and without (fit B) constant $a_0$. 

Fig. 3. clearly demonstrates the presence of both $c$-vortex and gluon components.
The gluonic contribution $\propto \kappa^4$ to the RG invariant condensate   
starts to dominate for $\kappa > \sqrt{\sigma}$.

\begin{figure}[!htb]
\vspace{-5mm}
\epsfxsize=6.6cm \epsffile{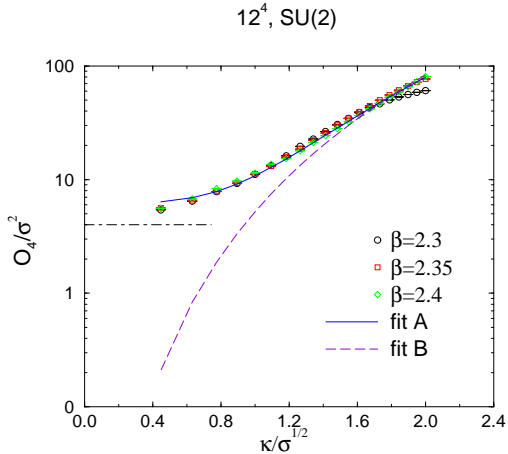}
\vspace{-12mm}
\caption
{$O_4/\sigma^2$ vs. cooling scale $\kappa/\sqrt{\sigma}$}
\label{fig:3}
\end{figure}
\vspace{-6mm}

The constant $a_0$ specifies the ultimate $Z(2)$ vortex content consistent with 
$O_4 = \lim_{\kappa \to 0}~O_4(\kappa) \approx 0.10 \dots 0.15~\mathrm{GeV}^4$
which is in the ballpark of recent $SU(2)$ gluon condensate estimates.
After Landau gauge fixing the density $\rho$ of defect links with $Z_{x,\mu}=-1$ 
has been also found to be RG invariant, 
$\rho \approx 0.7~\mathrm{fm}^{-4}$\cite{KL-talk}. 

\section{$c$-VORTEX DOMINANCE OF THE GLUEBALL SPECTRUM}

Pure gluodynamics is characterized by a wide gap between the correlation length 
corresponding to the lowest gauge invariant excitations (glueballs) and the 
confinement scale of few $10^2$ MeV. 
In view of the proposed coset cooling method it is interesting to 
ask whether removing the coset, i.e. gluon, part  distorts the glueball 
spectrum. 
This question has been answered by a comparison of the lowest glueball states 
in uncooled and coset-cooled gauge field onfigurations reported in 
Ref.\cite{Schaefke}. For $\beta=2.3$, $2.4$ and $2.5$ on a $8^3\times16$ lattice 
the lowest $O^{+}$ glueball state (with mass $m_{O^{+}} = 1.67(11)$ GeV) 
and the lowest $2^{+}$ glueball state ($m_{2^{+}} = 2.30(8)$ GeV), 
have been identified first on uncooled configurations. Fig. 4 demonstrates how
the renormalized $O^{+}$ glueball correlator measured on coset-cooled
configurations is fitted by the exponential slope obtained from uncooled 
$SU(2)$ correlator data. Similarly, the $2^{+}$ glueball mass is reproduced.

\begin{figure}[!htb]
\vspace{-5mm}
\epsfxsize=6.6cm \epsffile{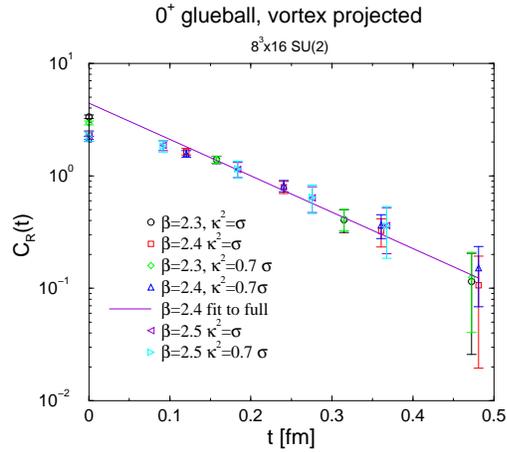}
\vspace{-12mm}
\caption
{Glueball propagator for coset-cooled configurations 
compared with the fit describing the glueball for uncooled configurations}
\label{fig:4}
\end{figure}
\vspace{-6mm}

\end{document}